\documentclass[aps,prb,twocolumn,superscriptaddress,showpacs]{revtex4-1}

\usepackage{graphicx}
\usepackage{subfig}
\usepackage{amsmath}
\usepackage{bm}
\let\oldhat\hat
\renewcommand{\vec}[1]{\bm{#1}}
\renewcommand{\hat}[1]{\oldhat{\bm{#1}}}
\newcommand{\tens}[1]{\underline{\underline{\bm{#1}}}}
\usepackage{hyperref}

\begin{document}

\title{Multiple light scattering and near-field effects in a fractal tree-like ensemble of dielectric nanoparticles}
\author{Matteo Gerosa}
\email[To whom correspondence should be addressed. Electronic address: ]{matteo.gerosa@polimi.it}
\affiliation{Department of Energy, Politecnico di Milano, via Ponzio 34/3, 20133 Milano, Italy}
\author{Carlo Enrico Bottani}
\affiliation{Department of Energy, Politecnico di Milano, via Ponzio 34/3, 20133 Milano, Italy}
\affiliation{Center for Nano Science and Technology @Polimi, Istituto Italiano di Tecnologia, via Pascoli 70/3, 20133 Milano, Italy}

\date{\today}

\begin{abstract}
We numerically study light scattering and absorption in self-similar aggregates of dielectric nanoparticles, as generated by simulated ballistic deposition upon a surface starting from a single seed particle. The resulting structure exhibits a complex tree-like shape, intended to mimic the morphologic properties of building blocks of real nanostructured thin films produced by means of fine controlled physical deposition processes employed in nanotechnology. The relationship of scattering and absorption cross sections to morphology is investigated within a computational scheme which thoroughly takes into account both multiple scattering and near-field effects. Numerical results are compared with a pre-existing single scattering limited analytical treatment of light scattering in fractal aggregates of small dielectric particles.
\end{abstract}


\maketitle

\section{Introduction}

Light scattering off aggregates of small dielectric particles has been extensively studied in the past with the aim of understanding the optical properties of a wide variety of physical systems. Typical applications are found in the fields of astrophysics, atmospheric and biological physics (see for instance Ref.~\onlinecite{ishimaru} for a broad review and Refs.~\onlinecite{Brunsting1972}, \onlinecite{Purcell1973}, \onlinecite{Berry1986} for selected examples of application to specific problems). In more recently published works, some of the previously developed computational techniques have been exploited in order to investigate the optical behavior of small-sized assemblies of nanoparticles. In Ref.~\onlinecite{Girard1990} a model to theoretically describe the optical interaction between a dielectric tip and a nanostructured corrugated surface in optical microscopy experiments (SNOM) is proposed. In Ref.~\onlinecite{Girard2005} a detailed summary of the principal theoretical approaches to nano-optics is presented, mainly oriented to the interpretation of optical images of nanostructures.

In this paper we numerically investigate light scattering and absorption in porous agglomerates of nanometer-sized dielectric particles obtained by simulated ballistic deposition on a flat surface substrate. The model is intended to understand light propagation throughout a structurally discrete medium whose morphologic properties mimic that of hierarchically-organized nanostructured thin films produced by means of physical deposition techniques.\cite{DiFonzo2009} To name one potential application, it is believed that such materials could play a pivotal role in the development of highly efficient photoanodes for novel photovoltaic devices (dye-sensitized solar cells, DSSCs).\cite{Sauvage2010}

In the present work much of the effort will be put into shedding light on the relationship between integral parameters describing the overall optical behavior of the system and its morphologic properties. Our results are put in comparison with the predictions of the analytical scalar theory developed by Berry and Percival\cite{Berry1986} concerning the optics of a generic fractal aggregate of tiny particles, which nevertheless completely neglects both multiple scattering effects and the vector nature of light which, instead, we thoroughly take into account. 
Our calculations are based on an ad hoc modified version of the well-known coupled dipole approximation,\cite{Draine1994}$^{,}$\cite{Mulholland1994}$^{,}$\cite{Yurkin2007} first introduced by Purcell and Pennypacker in 1973,\cite{Purcell1973} in relation to the investigation of the scattering properties of arbitrarily shaped dielectric particles. In the first section we critically review the main concepts underlying the coupled dipole approach and discuss its physical interpretation introducing a new modified version of it. The rest of the paper is fully devoted to the investigation of the optical properties of fractal aggregates of nanoparticles obtained by ballistic deposition upon a surface, starting from a single seed particle. A systematic analysis is conducted of how the morphologic characteristics of the scattering medium are able to shape its optical properties. It is found out that, besides ``macroscopic'' features such as the total number of constituent particles, also the local properties of matter distribution is essential in this regard. Finally, the importance of multiple scattering effects is emphasized both for conceptual reasons and in view of gaining a deep understanding of the relationship of optics to morphology, as well as that between scattering and absorption processes, which is closely connected to the issue of enhancing absorption by relying on effective light trapping\cite{Wiersma1997} in real devices, such as DSSCs.\cite{Ferber1998}  Both qualitative and quantitative differences are shown to exist with the results of the single scattering limited theory by Berry and Percival.

\section{Coupled dipole model re-examined}

We begin by considering a system composed by $N$ identical dielectric spheres with radius $a\ll\lambda$, where $\lambda$ is the wavelength of the incident radiation, assumed to be a linearly polarized monochromatic plane wave. We suppose each particle to be an optically homogeneous and isotropic body, being fully characterized by the appropriate dielectric constant $\epsilon(\omega)$ of the constituent material, evaluated at the frequency $\omega$ of the incoming wave. Since, by assumption, the external electric field varies slowly over the region occupied by a single particle, it is allowed to apply the Rayleigh approximation for scattering. Within this approximation, we can assume –--to lowest order--– that the field scattered by each nanoparticle is due to the electric dipole moment induced in it by the external wave only. The latter assumption implies no interaction at all between distinct dipoles: each particle radiates independently as though no other one is present in the vicinity, according to the so-called single scattering approximation (also known as Born or Rayleigh-Debye scattering approximation in other contexts).\cite{ishimaru}

If the particles are close enough to each other (e.g. in a topologically connected ensemble of particles), we expect, in principle, that the effects of mutual interactions between dipoles are not negligible and should be taken into account in the description. This is made possible by a perturbative approach in which the resulting scattered field and the discrete dipole moment distribution induced within the scattering structure are computed self-consistently. According to this picture, the first order electric field impinging on the $j$-th particle is given by
\begin{equation}
\vec{E}^{(1)}(j)=\vec{E}^{(0)}(j)+\sum_{j'\neq j}\underline{\underline{\vec{P}}}(j,j')\cdot\vec{E}^{(0)}(j'),
\label{eq:firstorder}
\end{equation}
where $\vec{E}^{(0)}(j)=\vec{E}_0\,e^{i\vec{k}\cdot\vec{r}_j}$ is the $\omega$-component of the incident electric field ---having constant amplitude $\vec{E}_0$--- evaluated at the center of the $j$-th particle (the vacuum dispersion relation $\omega=c\left|\vec{k}\right|$ is assumed here, since the wave propagates in free space until it interacts ``pointwisely'' with the structure). The dipole moment propagator\cite{Jackson} (CGS units)
\begin{widetext}
\begin{equation}
\tens{P}(j,j')=V\alpha_{\omega}\left[k^2 \left(\tens{I}-\vec{n}(j,j')\vec{n}(j,j')\right)+\left(3\vec{n}(j,j')\vec{n}(j,j')-\tens{I}\right)\left(\frac{1}{r(j,j')^2}-\frac{ik}{r(j,j')}\right)\right]\frac{e^{ikr(j,j')}}{r(j,j')}
\label{eq:propagator}
\end{equation}
\end{widetext}
is a second-rank tensor describing correlation effects between dipoles and fully including also near-field effects. Here $V=(4\pi/3)a^3$ is the particle volume, $\alpha_{\omega}$ its polarizability per unit volume, $r(j,j')=\left|\vec{r}(j,j')\right|=\left|\vec{r}_j-\vec{r}_{j'}\right|$ the distance between the centers of $j$-th and $j'$-th particles, $\vec{n}(j,j')=\vec{r}(j,j')/r(j,j')$ the unit vector identifying the scattered field propagation direction and $\tens{I}$ the identity tensor. The last term on the right-hand side of Eq.~\eqref{eq:firstorder} represents the total zeroth order scattered field acting on the $j$-th particle (where the ``generalized dot product'' between a second rank tensor and a vector is defined, in Cartesian components, as $\tens{T}\cdot\vec{v}\equiv\sum_{\nu}T_{\mu \nu}v_{\nu}$). 

We now may wonder whether $\vec{E}^{(1)}(j)$ is still smooth enough to allow for Rayleigh theory to be employed. Indeed, as shown by the appearance of the propagator \eqref{eq:propagator}, the near-field contribution to the scattered field is strongly non-uniform in the proximity of the particle surface, due mainly to the $1/r^3$ term. This suggests that ---beyond the single scattering approximation and particularly in high-density systems--- the field experimented by a given particle as a result of the waves scattered by its neighbors at lower order may vary appreciably over the region occupied by it. As a consequence, the central assumption on which all Rayleigh theory is built –--that the exciting electric field is weakly dependent on space--– is not in principle satisfied by higher order scattered field components. Hence, one might point out that a description going beyond the electric dipole approximation is needed, relying on the rigorous results of Mie theory for scattering by arbitrarily sized dielectric spheres. On the other hand, it is easy to guess that such a detailed treatment would result in an explosion of the computation time, thus severely limiting the maximum dimension of the system (i.e. total number of constituent particles) able to be examined.

One way to take account of the field space fluctuations without abandoning the electric dipole scheme is to introduce for each particle –--and at scattering orders $n$ higher than zero--– an effective exciting field defined as
\begin{equation*}
\vec{E}^{(n)}(j)=\frac{1}{V} \int_{V}\vec{E}^{(n)}(\vec{r}_j)\,d^3r_j
\end{equation*}
i.e. the total $n$-th order electric field averaged over the single particle volume. We are then allowed to retrieve the Rayleigh approximation by applying it to the smoothed out electric fields $\left\{\vec{E}^{(n)}(j)\right\}_{j=1,\dots,N}$, giving rise to the $n$-th order effective dipole distribution $\left\{\vec{p}^{(n)}(j)\right\}_{j=1,\dots,N}$ through the quasi-static formula
\begin{equation*}
\vec{p}^{(n)}(j)=a^3\left(\frac{\epsilon(\omega)-1}{\epsilon(\omega)+2}\right)\vec{E}^{(n)}(j).
\end{equation*}
Equation~\eqref{eq:firstorder} can easily be generalized by recursion to yield the ``exact'' (to within the electric dipole scheme) self-consistent field at point $\vec{r}_a$ in free space:
\begin{widetext}
\begin{equation}
\vec{E}(a)=\vec{E}^{(0)}(a)+\sum_{j=1}^{N}\tens{P}(a,j)\cdot\vec{E}^{(0)}(j)+\sum_{j=1}^{N}\tens{P}(a,j)\cdot\left[\sum_{j'\neq j}\tens{P}(j,j')\cdot\vec{E}^{(0)}(j')\right]+\dots
\label{eq:wholeseries}
\end{equation}
\end{widetext}
where $\tens{P}(\cdot,j)$ is re-defined to be the propagator for the scattered field emitted by the effective dipole moment induced in the $j$-th sphere.

The infinite series in Eq.~\eqref{eq:wholeseries} is straightforwardly interpreted as the mathematical expression of the presence of multiple scattering processes switching on correlations between dipoles. It turns out that such processes can be intuitively visualized through a diagrammatic representation of Eq.~\eqref{eq:wholeseries}, which reveals many similarities with Feynman diagrams describing the motion of a single quantum particle in an external force field. Nonetheless, this resemblance can hardly be exploited to yield an analytical closed form expression for the total field. In analogy with the quantum case, the series~\eqref{eq:wholeseries} is indeed impossible to be evaluated, if not when the form of the particle distribution function is particularly simple (e.g. in the case of a completely disordered medium).

On the contrary, when the scattering medium is endowed with non trivial morphologies, it becomes necessary to attack the problem numerically, by computing the self-consistent field iteratively.

\begin{figure}[tb]
\centering
\includegraphics[width=1\columnwidth]{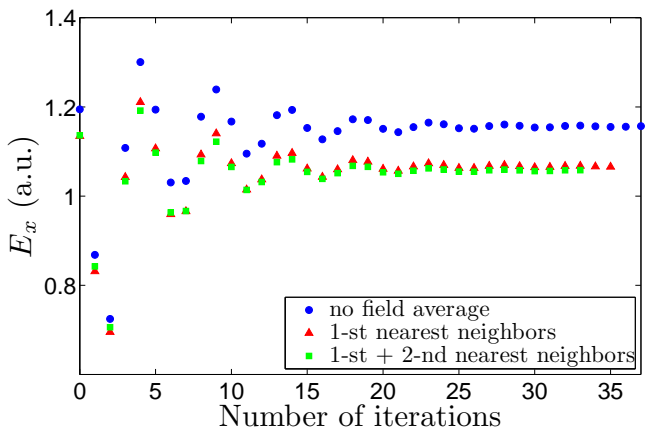}
\caption{\label{fig:fig1}Iterative calculation of the effective electric field (polarized component) acting on a chosen particle within an ensemble, showing the difference between the case in which the field average is not computed (blue dots) and the cases in which it is performed over the fields scattered by the first (red triangles) and jointly by the first and second nearest neighbor particles (green squares). The particle under consideration is located at the center of a cubic closed packed structure filled with spheres of radius $a=10\,\text{nm}$. The external field is an $x$-polarized monochromatic plane wave of wavelength $\lambda=100\,\text{nm}$ and unit amplitude.}
\end{figure}

Figure~\ref{fig:fig1} shows the typical behavior of the effective field acting upon a given particle as iteration proceeds. It also compares the case in which no field average is performed with the case in which it is, either over the field scattered by the first nearest neighbor particles or jointly by the first and second nearest neighbors. We note that the difference in the final effective field is appreciable, particularly between the two former situations. The effect of field smoothing vanishes as the second nearest neighbor particles are also involved in the field average calculation, confirming that field non-uniformity is of some importance only in a confined region in the vicinity of the particle surface. It is worth noting that these discrepancies are indeed perceptible only on length scales comparable to the particle size, where near-field effects become predominant. In contrast, it has been observed that they do negligibly influence the value of global optical parameters such as integral scattering and absorption cross sections, which instead are determined only by the far field components of the total scattered field. In addition, the ratio $a/\lambda$ also plays a crucial role in determining quantitatively the relevance of such effects, which become increasingly significant as the wavelength is decreased at fixed particle size.

As shown in Fig.~\ref{fig:fig1}, the introduction of field smoothing can also imply some reduction in amplitude oscillations while reaching the final solution. In this respect, it is however of much more relative importance to introduce some additional damping mechanism which could help eliminating stable oscillations around the final solution and preventing from divergences. The following replacement is a standard way to do it (see for instance Ref.~\onlinecite{Purcell1973}, where it was employed for the first time within the coupled dipole approximation):
\begin{equation*}
\vec{E}^{(n)}(j) \longmapsto \eta\,\vec{E}^{(n)}(j)+(1-\eta)\,\vec{E}^{(n-1)}(j)\qquad j=1,\dots,N,
\end{equation*}
where $0<\eta\le1$ must be fixed a priori (usually, $\eta=0.5$ is a proper choice).

\begin{figure}[tb]
\centering
\subfloat[]
{\includegraphics[width=1\columnwidth]{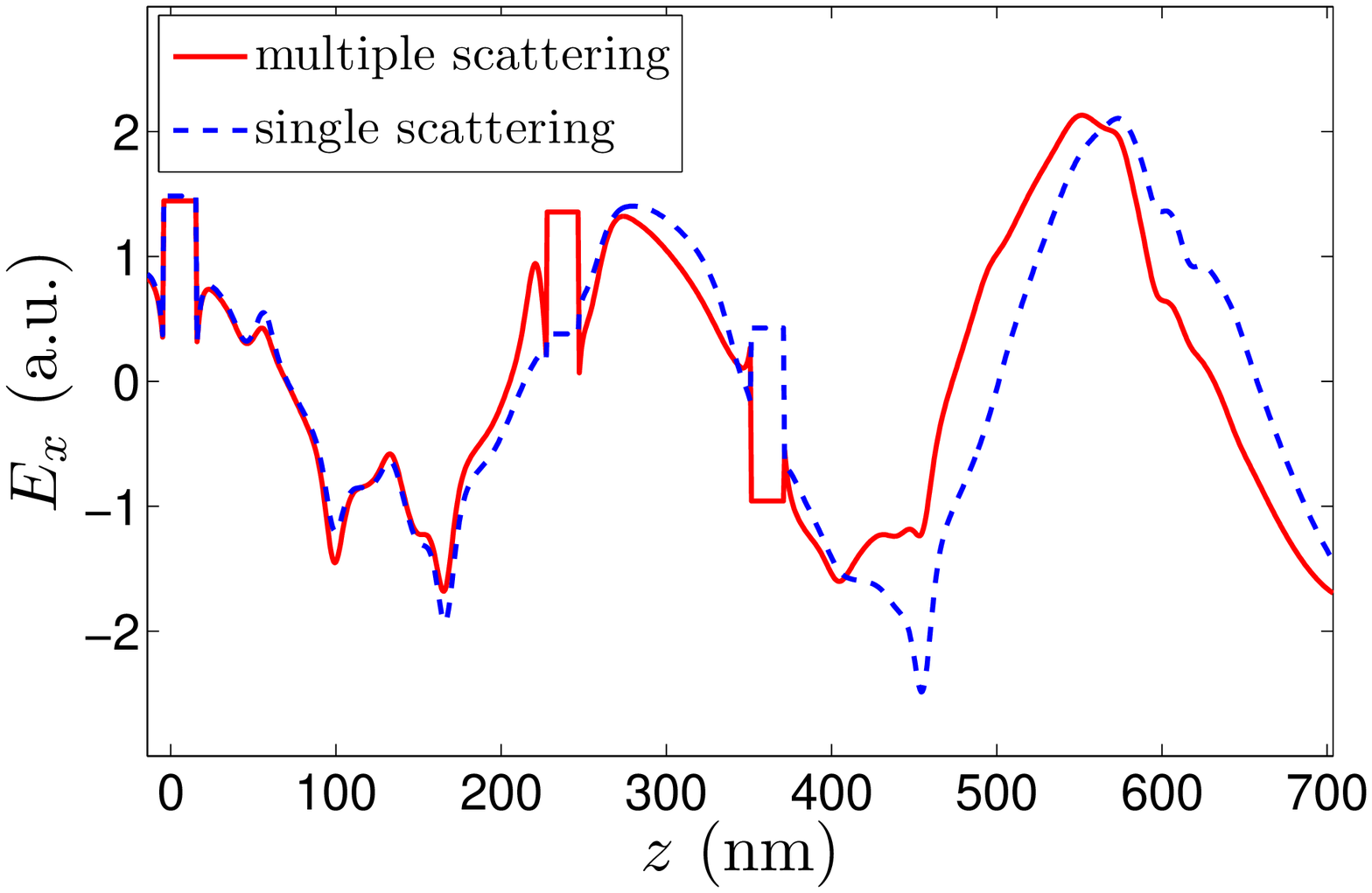}}\\
\subfloat[]
{\includegraphics[width=1\columnwidth]{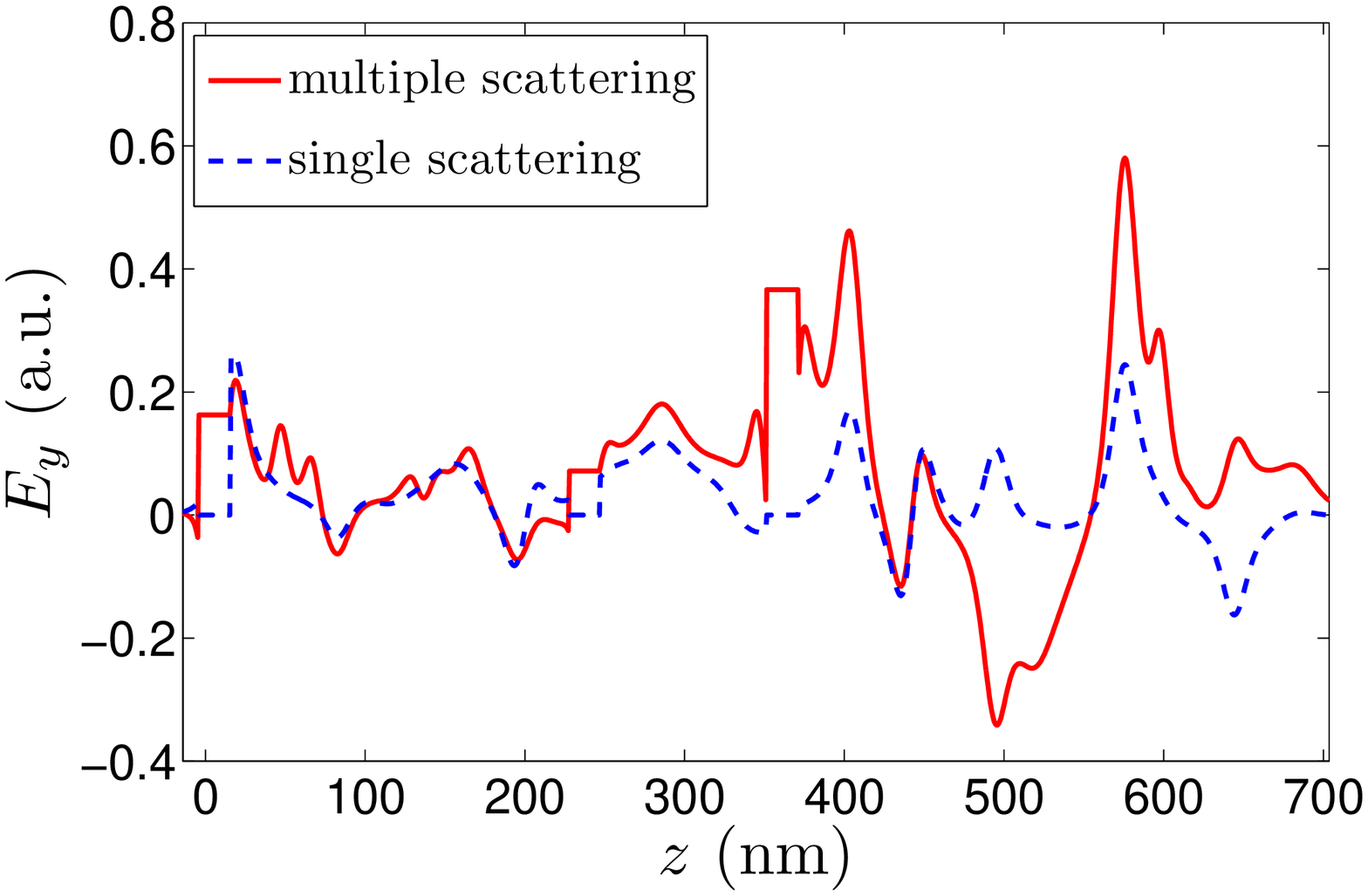}}\\
\caption{\label{fig:fig2}Electric field distribution along the axis of the tree-like structure represented in Fig.~\ref{fig:fig3}(b) ($z$ direction), showing the difference between the results of single and multiple scattering calculations. (a) Polarized and (b) depolarized electric field component along $y$. The incident plane wave has unit amplitude, propagates along $z$ and is linearly polarized along $x$.}
\end{figure}
One can wonder about the ultimate relevancy of including the description of multiple scattering in the model. There are different good reasons to convince oneself it is. To start with, as shown in Fig.~\ref{fig:fig2}, field spatial distribution differs both qualitatively and quantitatively whether multiple scattering is fully (although at finite order) taken into account or not. Such difference becomes even more appreciable in the depolarized field components, which sounds reasonable once the interaction between dipoles is understood as the original cause of depolarization of electric dipoles off the direction of the incident electric field. It is in fact common in the literature to attack the single scattering problem in the context of a much simpler scalar theory.\cite{Sorensen2001} Even more remarkably, it turns out that the standard definition of cross sections within the coupled dipole model is not compatible with the single scattering approximation. The expressions for extinction, absorption and scattering cross sections for an ensemble of electric dipoles are found in the paper by Purcell and Pennypacker\cite{Purcell1973} as well as in the works by Draine:\cite{Draine1994}$^{,}$\cite{Draine1988}
\begin{equation}
\begin{split}
&\sigma_{ext}=\frac{4\pi k}{E_0^2}\sum_{j=1}^{N} \Im\left\{\vec{E}^{(0)}(j)\,^{*}\cdot \vec{p}(j)\right\},\\
&\sigma_{abs}=\frac{4\pi k}{E_0^2}\frac{1}{V}\frac{\alpha_{\omega}''}{\left|\alpha_{\omega}\right|^2}\sum_{j=1}^{N}\left|\vec{p}(j)\right|^2,\\
&\sigma_{sca}=\sigma_{ext}-\sigma_{abs},
\end{split}
\label{eq:crosssections}
\end{equation}
where $E_0$ denotes the incident field amplitude, $k$ the wavevector norm and \[\alpha_{\omega}=\alpha'_{\omega}+i\alpha''_{\omega}=\frac{3}{4\pi}\left(\frac{\epsilon(\omega)-1}{\epsilon(\omega)+2}\right)\]
is the dielectric sphere polarizability per unit volume. It is easy to see by Eqs.~\eqref{eq:crosssections} that ---within the single scattering approximation and in the presence of a purely scattering medium (in which case $\alpha_{\omega}$ is real)--– the integral extinction cross section vanishes, which is not acceptable from the physical point of view. In other words, the forward-scattering theorem (from which expressions~\eqref{eq:crosssections} immediately follow) is not satisfied in the single scattering limit and therefore its application leads to biased results. We conclude by asserting the necessity of taking into account multiple scattering processes in order to infer properly about the optical behavior –--both in the far and in the near field region--- of the system.

To summarize, the coupled dipole model has been revisited in order to investigate the applicability of Rayleigh approximation in the presence of multiple scattering effects within a granular nanomaterial. A new effective field approach has been proposed beyond the single scattering approximation, in order to deal with possible field non-uniformities associated with near-field effects and to motivate rigorously the application of Rayleigh theory. Although in the cases we examined –--and for our purposes, i.e. the evaluation of integral cross sections--- the introduction of the effective field idea seems of little practical importance, we will keep using it throughout all the calculations whose results are discussed below, both for the sake of conceptual rigor and because in principle it may prove relevant in situations we did not consider and/or for purposes different from ours.

\begin{figure*}[tb]
\includegraphics[scale=.9]{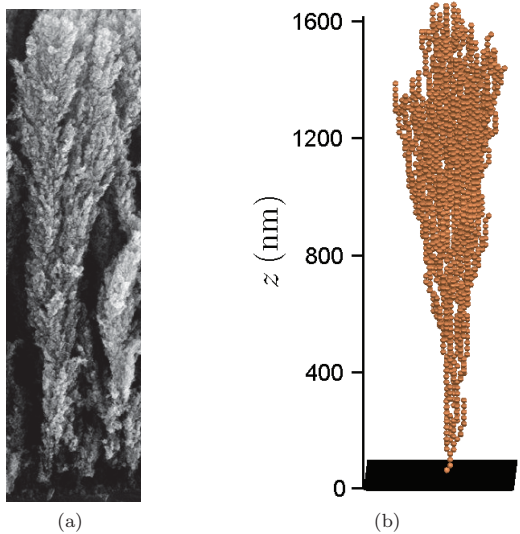}
\caption{\label{fig:fig3}(a) SEM image of a typical tree-like aggregate of semiconductor oxide nanoparticles as produced by pulsed laser deposition and (b) as obtained by simulating the ballistic deposition process starting from a single seed particle.}
\end{figure*}

\section{Optical properties dependence on morphology in fractal nanostructured media}
We now consider the application of the electromagnetic scattering model discussed above to the investigation of the optical properties of agglomerates of dielectric nanoparticles obtained by ballistic deposition on a flat surface. In particular, we are interested in elucidating the relationship between the global optical behavior and the structural characteristics of such systems.
In doing this, we will often make reference to the original work by Berry and Percival\cite{Berry1986} concerning the optics of smoke released into the atmosphere, in which for the first time the importance of the fractal nature of the scattering medium is fully brought out.
We first begin by giving a brief account of the main morphologic features of the systems we wish to investigate.

\subsection{Ballistic deposition of spherical particles onto a surface}

Surface growth and sedimentation phenomena are frequently encountered in many branches of applied science and several theoretical accounts have been proposed over the years in order to provide adequate descriptions of various processes observed in nature.\cite{Vicsek} Typically, the prominent features of a certain real process are translated into a model system which preserves in the description only the relevant degrees of freedom under investigation. Among the wide variety of models proposed, one of the simplest is the ballistic deposition (BD) model, first introduced independently by Vold\cite{Vold1959}  and Sutherland\cite{Sutherland1966} to study sedimentation phenomena of colloidal particles. It has been demonstrated\cite{Meakin1993}$^{,}$\cite{Halpin1995}$^{,}$\cite{Robledo2011} that the same description can be usefully adopted in accounting for the formation of morphologically complex thin films obtained as the by-product of nanoparticle deposition processes on a substrate, typically employed in several nanofabrication techniques. Recent experimental studies\cite{DiFonzo2009} using the PLD (Pulsed Laser Deposition) technique have proved that  ---provided the deposition parameters are accurately tuned--- it is possible to grow a hierarchically-organized nanostructured film exhibiting high porosity at the nano- and meso-scale. Figure~\ref{fig:fig3}(a) shows the typical appearance of the resulting material, which reveals multiscale structural organization involving length scales ranging from tens of nanometers to some micrometers. Each nanoparticle is readily identified as the elementary constituent of a complex forest-like structure arranged in several neighboring tree-like interacting aggregates, which endow the system with the typical porosity. Each tree can be conceived as growing starting from a single seed particle deposited on the supporting surface and evolving in time according to an appropriate stochastic rule. The essential prescription on which the BD model is based is that each particle rigidly sticks to the growing surface upon contact with another particle already deposited. In our model we have assumed that the growth process is parametrized by two distinct sticking probabilities, according to whether the depositing particle first comes into contact with an underlying one ($p_u$) or one displaced from its migration direction ($p_s$), which is chosen to be orthogonal to the supporting surface. Different ratios $p_u/p_s$ end up in different morphologies of the resulting structure.

In the following, we will limit our considerations to the case of single tree-like structures (Fig.~\ref{fig:fig3}(b)) made up of up to 2500 spherical nanoparticles having radius $a=10\,\text{nm}$. In order to provide a quantitative description of their structural features, we identify each aggregate with a pair of scalar parameters: the total number of constituent particles $N$ and its fractal dimension $D$. The latter is readily calculated once the position of each particle within the structure is known, according to the formula\cite{Berry1986}
\[
D=\frac{\log N}{\log(R/a)},
\]
where $R$ defines a characteristic linear dimension of the agglomerate, being proportional to its radius of gyration:
\[
R=\left[\frac{2}{N}\sum_{j=1}^{N}\left|\vec{r}_j-\vec{r}_{cm}\right|^2\right]^{1/2},
\]
$\vec{r}_j$ being the position vector of the center of the $j$-th particle and $\vec{r}_{cm}$ the center of mass position vector of the whole aggregate.
We will assume that such parameters are indeed sufficient to fully identify the salient morphologic properties of the system under consideration and significant enough to make explicit possible substantial connection with its optical features. The remaining part of this paper will be devoted to clarifying this last important point.

\subsection{Relationship of optics to morphology in tree-like fractal structures}

The issue of how the optical behavior of an agglomerate of small dielectric particles is influenced by the features of the particle distribution itself is of both fundamental and practical (application-oriented) interest. While in perfectly ordered arrangements of scatterers (such as an ideal atomic lattice) general symmetry considerations allow in principle to directly grasp the salient characteristics of the optical response of the system, it is not quite as an easy task to do the same for fully and, even more problematically, partially disordered ensembles of particles. However, it turns out that several real systems arising from processes of aggregation or sedimentation often exhibit –--at least in an approximate fashion--– the remarkable property of being invariant under scale transformations. Such self-similarity is fully accounted for by a single scalar parameter –--the fractal dimension of the aggregate. The very first attempt to establish the dependence of integral optical quantities (cross sections) on fractal dimension was made by Berry and Percival\cite{Berry1986}, who employed a single scattering limited analytical approach to investigate the optics of soot-like agglomerates of particles.

In order to better clarify the possible relevance of our analysis for emerging applications, we will study more specifically the typical configuration of a single tree structure in a dye-sensitized solar cell. In such devices, the presence of a monolayer of a suitable highly absorbing molecule coating each nanoparticle makes light absorption strongly localized in the proximity of the sensitized available surface of the structure (which is enhanced by the typically high porousness at the nanoscale). Instead, non-sensitized nanoparticles are typically made of a large band-gap semiconducting material (such as titanium dioxide), being thus transparent (i.e. purely scattering) to radiation in the frequency range of interest (which, in view of application to solar cells, is restricted to the visible frequency spectrum). Our model assumes each nanosphere to be coated by a thin purely absorbing layer characterized by an appropriate dielectric constant $\epsilon_{l}(\omega)=1+i\epsilon_l''(\omega)$ which, suitably combined with the bulk sphere real dielectric constant $\epsilon_b(\omega)$, results in the coated particle polarizability $\alpha_{\omega}$ (see Ref.~\onlinecite{Bohren} for details on the derivation) which is also dependent on the particle radius $a$ and the layer thickness $\delta$ (in our simulations, we assumed $a=10\,\text{nm}$, $\delta=0.1\,\text{nm}$, $\epsilon_l''(\omega)=3$, $\epsilon_b(\omega)=5.8$, which corresponds\cite{Asahi2000} to the purely real bulk dielectric constant of titanium dioxide at $\lambda=500\,\text{nm}$).

The basic simplifying assumption which underlies the analytical treatment by Berry and Percival is that multiple scattering effects can be completely neglected, provided a limiting condition on a suitable combination of $D$ and $N$ is satisfied. Intuitive considerations suggest that, when $D<2$, single scattering effects are predominant at all stages of aggregation (i.e. irrespective of the total number of constituent particles). On the contrary, for $D>2$, approximate estimates show that multiple scattering is not always negligible, becoming in fact important when $N\sim\widetilde{N}\equiv(ka)^{-D/(D-2)}$. If we strictly followed this prescription, we would not need in principle to take into account multiple scattering, being our systems typically characterized by fractal dimensions slightly larger than 2 and comprising at most some thousand particles (for $\lambda=500\,\text{nm}$, $D=2.1$, $a=10\,\text{nm}$, $\widetilde{N}$ exceeds $N$ by several orders of magnitude).

\begin{figure}[tb]
\includegraphics[width=1\columnwidth]{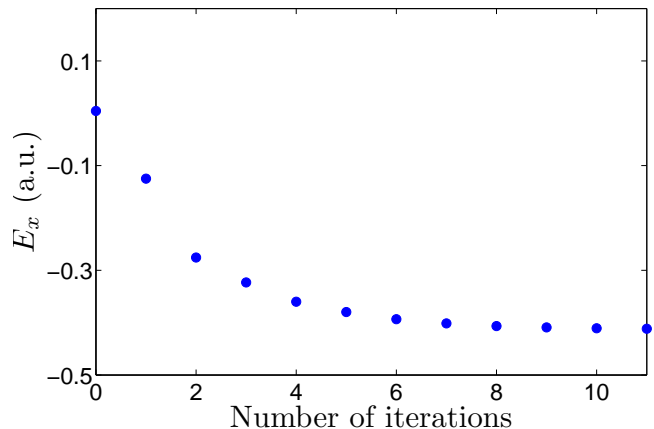}
\caption{\label{fig:fig4}Evolution of the effective electric field (polarized component) acting upon a chosen particle within a tree-like aggregate made up by 1000 particles (here, the zeroth iteration corresponds to the single scattering regime). The incident field is a plane wave with wavelength $\lambda=300\,\text{nm}$ and unit amplitude. The wavevector is directed along the tree axis.}
\end{figure}

However, some indications lead us to speculate that multiple scattering is not as a negligible effect as suggested by the above discussion. As shown in Fig.~\ref{fig:fig4}, the effective electric field experienced by a given particle within the ensemble may be significantly different in value whether it is obtained via the self-consistent calculation or not (i.e. within the single scattering approximation). We are then led to conclude that multiple scattering effects do in principle play a decisive role in the cases of our concern. Furthermore, we remind that the definitions of cross sections provided above are inapplicable whenever multiple scattering is neglected.

The results of the analytical treatment by Berry and Percival indicate that scattering and absorption cross sections for the whole aggregate are proportional to a power of the total number of constituent particles, the characteristic exponent being in general a function of the fractal dimension. It turns out that, for $D>2$,
\begin{equation}
\begin{split}
&\sigma_{sca}\propto N^{2-2/D},\\
&\sigma_{abs}\propto N,
\end{split}
\label{eq:powerlaws}
\end{equation}
while, for $D<2$, the scattering per particle becomes independent of $N$.

\begin{figure}[tb]
\centering
\subfloat[]
{\includegraphics[width=1\columnwidth]{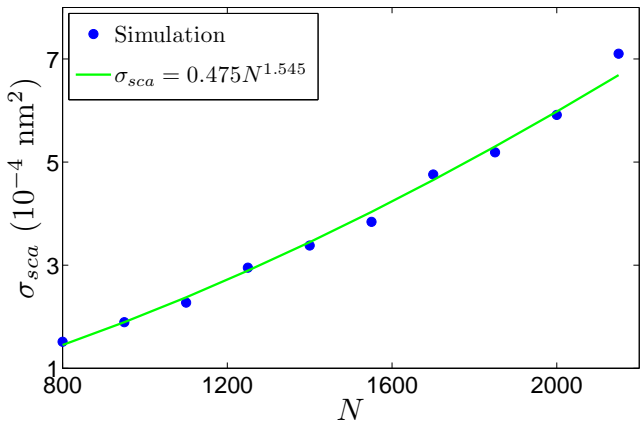}}\\
\subfloat[]
{\includegraphics[width=1\columnwidth]{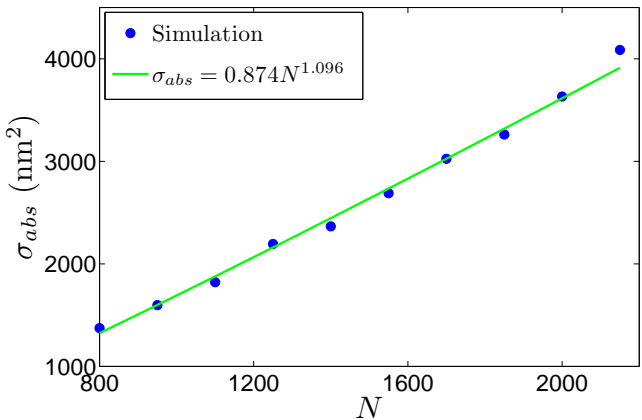}}
\caption{\label{fig:fig5}(a) Scattering and (b) absorption cross sections as a function of $N$ for tree-like aggregates obtained by ballistic deposition. The fractal dimension of each is kept fixed to within a small interval around $D=2.17$.}
\end{figure}
\begin{figure}[tb]
\centering
\includegraphics[width=1\columnwidth]{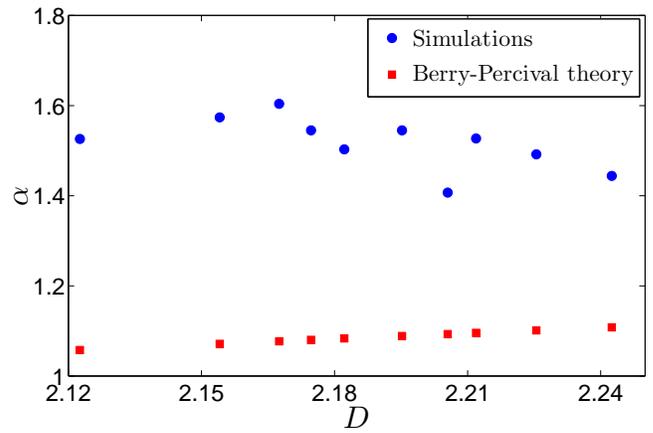}
\caption{\label{fig:fig6}Scattering exponent (defined through $\sigma_{sca}\propto N^{\alpha}$) as a function of the fractal dimension, as obtained by interpolating data from simulations carried out as shown in Fig.~\ref{fig:fig5} for different fractal dimensions. Comparison with the theoretical predictions by Berry and Percival.}
\end{figure}

Figure~\ref{fig:fig5} shows the results of our calculations for tree-like structures comprising an increasing number of particles and characterized by approximately the same fractal dimension. It is apparent that –--even when multiple scattering effects are fully taken into account--- the power law dependence of cross sections implied by theory is indeed respected. Actually the agreement is only qualitative in nature, since the values of the characteristic power law exponents turn out to be inconsistent with the analytical predictions. In particular, in all cases we examined (in which the fractal dimension roughly ranges from 2.1 to 2.25 and $N$ varies between 850 and 2150), we observed that theory systematically underestimates the growth rate of cross sections with increasing the size of the system. This effect is particularly significant as far as the scattering cross section is concerned, as clearly indicated by Fig.~\ref{fig:fig6}. It is reasonable to attribute such disagreement to the relevance of multiple scattering effects, which, being the result of dipole-dipole interactions, are indeed strongly dependent ---and in a more dramatic way with respect to the single scattering case--- on the number of particles constituting the system.

\begin{figure}[tb]
\centering{
\includegraphics[width=1\columnwidth]{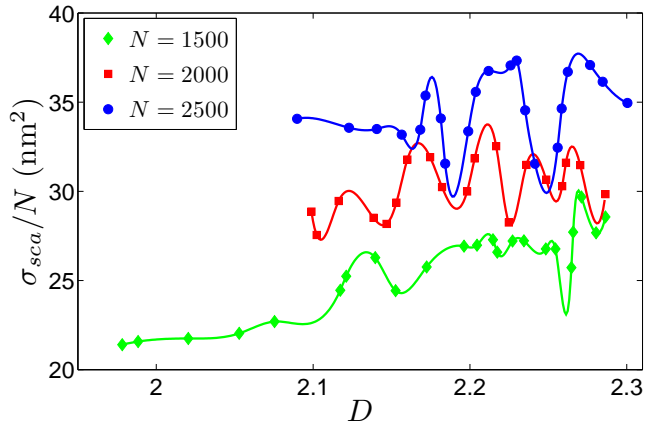}}
\caption{\label{fig:fig7}Scattering cross section per particle as a function of the fractal dimension, for three different system sizes ($N$).}
\end{figure}
The same conclusion is inferred when looking at the relationship between scattering cross section and fractal dimension, while keeping the system size fixed (Fig.~\ref{fig:fig7}). Here we find an additional evidence of the non-trivial scattering behavior of the medium when its fractal dimension significantly exceeds the value $D=2$ (which, even intuitively, represents a topological demarcation line between single and multiple scattering regimes). In fact, in the region far from that threshold, strong fluctuations of cross section are observed, something which is not consistent with the plain power law behavior implied by~\eqref{eq:powerlaws}. By contrast, as soon as the fractal dimension approaches the value $D=2$, oscillations are smoothed out and one recovers the correspondence with the analytical predictions, which suggest that the cross section per particle becomes independent of $N$ for particle distributions morphologically close to a surface.

Note that the results shown in Fig.~\ref{fig:fig7} refer to one single particle configuration for each value of the pair $(D,N)$. While, due to the stochastic nature of the ballistic growth process, fixing such quantities does not univocally determine the resulting configuration, it has been observed that two different situations can occur. One in which to a fixed $(N,D)$ there correspond morphologically similar configurations (whose fingerprint is provided by an additional morphologic parameter, the opening angle of the cone which fully contains the tree-like structure) which are characterized by cross section values determined to within an error of one percent. The other one (which has been observed to be much more rare than the previous one) in which, to the same $(N,D)$ considered above, there may correspond appreciably different configurations which are characterized by cross section values that can be remarkably different with respect to the ones observed in the previous case (the error can become of the order of 10 percent). It is worth noting that, in either cases, the fluctuations due to statistics are not such to be comparable to the typical size of the observed oscillations of the cross section as a function of the fractal dimension. Moreover, if for a fixed $(N,D)$ an average was performed over all the possible particle configurations, the rare ones would not weight significantly in determining the average cross section and, as such, would leave the situation in Fig.~\ref{fig:fig7} unchanged.

\begin{figure}[tb]
\centering{
\includegraphics[width=1\columnwidth]{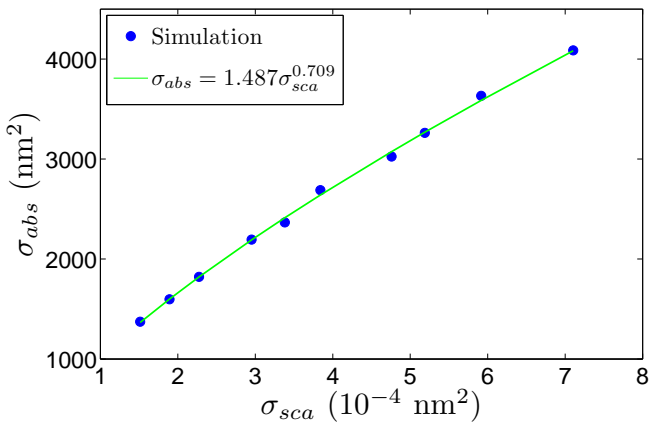}}
\caption{\label{fig:fig8}Absorption cross section as a function of the scattering cross section, which increases as a result of the increased number of particles constituting the system. The fractal dimension is kept fixed to within a small interval around $D=2.17$ and $N$ is let to vary from 850 to 2150 particles.}
\end{figure}

\begin{figure}[tb]
\centering{
\includegraphics[width=1\columnwidth]{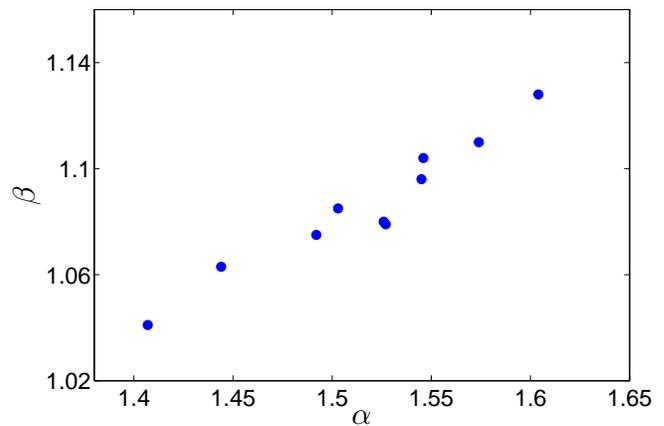}}
\caption{\label{fig:fig9} Absorption exponent (defined through $\sigma_{abs}\propto N^{\beta}$) as a function of the scattering exponent $\alpha$. Both depend exclusively on the fractal dimension of the structure.}
\end{figure}
Finally, we analyzed the relationship between scattering and absorption, the understanding of which is also motivated in view of future applications. If we limit our attention to aggregates exhibiting the very same self-similarity properties (i.e. having the same fractal dimension), it seems reasonable, even almost obvious, to guess that absorption is enhanced upon increasing the total number of particles (Fig.~\ref{fig:fig8}). Indeed, as soon as multiple scattering becomes increasingly strong as consequence of the growing number of dipole-dipole interactions, also absorption is intensified, being the result of an increased probability for the electromagnetic wave to multiply interact within the structure and, eventually, be absorbed.

It is perhaps less intuitive and understandable that also the small scale morphologic properties of the medium can play a role in determining its optical response. As a matter of fact, Fig.~\ref{fig:fig9} shows that this circumstance is verified. Since in the multiple scattering regime the absorption power law becomes characterized by a $D$-dependent exponent, it is interesting to look at its relationship with the scattering exponent, which is also a function of the fractal dimension only. It is observed that $\beta$ increases approximately linearly with $\alpha$, which can be easily interpreted as follows. As soon as the total number of particles within the system is kept fixed, the scattering cross section is allowed to vary only as a consequence of a change in the local morphologic configuration of the particle distribution, which, in our framework, is assumed to be fully described by the fractal dimension. Figure~\ref{fig:fig9} then implies that, whenever such a rearrangement causes the scattering exponent to increase, also the absorption exponent and, hence, cross section are enhanced. It is important to stress that such a peculiar behavior could not be inferred from the single scattering limited analysis carried out by Berry and Percival. Indeed, Eqs.~\eqref{eq:powerlaws} implies that $\beta=1$ irrespective of the fractal dimension. In fact, only taking into account how different particles mutually interact detailed information about the morphology of the system at any length scale is possible to be extracted. This latter observation serves as an additional hint of the non-trivial contribution of multiple scattering to the optical properties of nanostructures, which can eventually lead to the emergence of completely new behaviors (e.g. the absorption exponent $\beta$ becoming dependent on the fractal dimension) and/or to profound modifications of the ones already arising within the single scattering scheme.

Finally, the importance of the small scale morphologic properties of the structure is also made apparent in Fig.~\ref{fig:fig7}. The strong scattering oscillations with changing fractal dimension, appearing in the region far enough from $D=2$, due to a small change in the fractal dimension are shown to be quantitatively comparable to a variation in the scattering cross section resulting from a substantial change in the system size $N$ at fixed fractal dimension.

\section{Conclusions}

We have carried out an in-depth analysis of the optical properties of partially disordered self-similar aggregates of small dielectric particles, as obtained by performing a particle-by-particle ballistic deposition starting from a single seed placed on a flat surface. The deposition model being parametrized by two distinct sticking probabilities has allowed to gain control on the local morphology of the system, making it possible to explore a wide range of different particle configurations. This resulted in the capability to systematically study the relationship between cross sections and structural parameters, such as the total number of particles and the fractal dimension.

Such task has been achieved by exploiting a modified version of the coupled dipole model to describe the propagation of an electromagnetic plane wave throughout the structure.

After having observed how multiple scattering could be decisive in determining field distribution down to the smallest length scales, its influence has also been studied on the overall optical properties of the aggregate and the resulting relationship with its morphology. In particular, we have recognized that it too determines the emergence of qualitatively new cross section behaviors dominated by the local morphology of the system, something which could not be predicted within the framework of the approximate theory proposed by Berry and Percival. We also observed that multiple scattering is able to quantitatively modify the cross sections dependence on macroscopic structural parameters such as the total number of constituent particles.

Although our observations are unlikely fully representative of the complex optical behavior of a real hierarchical nanostructure, the present work may provide useful indications which should be taken into account in view of future further investigation, the feasibility of which has been mainly limited, in the present framework, by excessive computation load. The central message is that, generally speaking, the optical properties of a nanostructured material are non-trivially shaped by its inherent morphology, starting from the ``macroscopic'' down to the sub-micrometic scale. In this regard, multiple scattering and near-field effects are of crucial importance in revealing the underlying relationships.

One could be interested in understanding how light propagates in a structure made up by several tree-like aggregates of nanoparticles of the same kind considered above. In such a situation, the optical interaction among distinct trees brings in additional degrees of freedom to be considered, also in connection with the increased complexity of the resulting morphology. The consequent optical behavior may be altered also perhaps from the qualitative point of view, possibly due to the emergence of remarkable diffraction effects. Such analysis would obviously be of paramount importance also in consideration of technological applications, providing a fundamental description of complex optical phenomena such as light trapping in advanced photovoltaic cells.\cite{Zhang2012}

\begin{acknowledgments}
We are thankful to C.S. Casari and A. Li Bassi for providing the SEM image in Fig.~\ref{fig:fig3}.
\end{acknowledgments}

\bibliography{references}
\end{document}